\documentclass[aps,prl,twocolumn,superscriptaddress]{revtex4-2}
\usepackage{epsfig}
\usepackage{graphicx}
\usepackage[figuresright]{rotating}
\usepackage{amsmath}
\usepackage{amsfonts}
\usepackage{amssymb}
\usepackage{mathrsfs}
\usepackage{dcolumn}
\usepackage{bm}
\usepackage{color}
\usepackage[
    colorlinks = true,
    linkcolor = blue,
    urlcolor  = blue,
    citecolor = blue,
    anchorcolor = blue
]{hyperref}

\def\be{\begin{equation}} \def\ee{\end{equation}}
\def\bea{\begin{eqnarray}} \def\eea{\end{eqnarray}}

\newcommand{\GrantNo}[1]{No.\,#1}
\newcommand{\secheader}[1]{\textit{#1\,---}}
\newcommand{\FIG}[1]{Fig.\,\ref{#1}}
\newcommand{\EQ}[1]{Eq.\,(\ref{#1})}

\newcommand{\WQCASQC} { Wilczek Quantum Center and Key Laboratory of Artificial Structures and Quantum Control, School of Physics and Astronomy, Shanghai Jiao Tong University, Shanghai 200240, China}

\newcommand{\tdli}{T. D. Lee Institute, Shanghai Jiao Tong University and Shanghai Research Center for Quantum Sciences, Shanghai, China}

\begin{document}
\title{Infinite-temperature quantum phases and phase transitions}

\author{Tengzhou Zhang}
\affiliation{\tdli}
\affiliation{\WQCASQC}

\author{Zhizhen Chen}
\affiliation{\tdli}
\affiliation{\WQCASQC}

\author{Zi Cai}
\email{zcai@sjtu.edu.cn}
\affiliation{\WQCASQC}

\begin{abstract}  

In this study, we reveal nontrivial quantum physics in an infinite-temperature system. By performing an unbiased quantum Monte Carlo simulation, we study a hybrid model composed of hard-core bosons, whose hopping amplitude is mediated by the density of another type of soft-core bond bosons that can absorb entropy indefinitely. It is shown that the Bose-Einstein condensate can persist in three dimensions even when the temperature approaches the infinite-temperature limit. In contrast, in two dimensions, the quasi-superfluid is depleted by the fluctuations of the bond bosons, which, on the other hand, enhance the conductivity of the hard-core bosons in the normal phase. A generalization to the fermionic model has also been discussed.

\end{abstract}


\maketitle
\secheader{Introduction} The search for high-temperature macroscopic quantum coherent states, such as room-temperature superconductivity, has been a major focus of quantum physics for decades. However, it is widely expected that thermal fluctuations at high temperatures disrupt quantum interference, leading to a featureless mixed state where quantum entanglement and coherence are suppressed. Consider the infinite-temperature limit as an example: for a lattice model with a finite Hilbert space dimension per site, an infinite-temperature state is characterized by a unit density matrix with no quantum coherence or correlation. Despite this, the relaxation dynamics \cite{Pichler2010,Marino2012,Poletti2012,Cai2013,Sieberer2013,Poletti2013,Buchhold2015,Cai2020,Cai2022} and transport behavior \cite{Ljubotina2019,Nardis2019,Dupont2019} around this featureless state can still be highly nontrivial and universal. However, for a quantum many-body state with an unbounded local Hilbert space, the situation could be qualitatively different, opening up new possibilities to explore nontrivial quantum physics near the infinite-temperature state.

A related phenomenon is spontaneous symmetry breaking, which is traditionally thought to occur only at low temperatures but may persist at arbitrarily high temperatures. Weinberg developed a quantum field model with heat-resistant order at intermediate temperatures \cite{Weinberg1974}, and Chai {\it et al.} constructed an ultraviolet-complete theory based on conformal field theory, which could exhibit order at any temperature \cite{Chai2020,Chai2020b}. More recently, a general mechanism has been proposed to account for high-temperature order in a variety of models \cite{Han2025}, where the systems of interest couple to bosonic baths that can absorb entropy indefinitely, thus maintaining the system in a low-entropy ordered phase. This is reminiscent of the Pomeranchuk effect in $^3$He \cite{Pomeranchuk1950}, where the spin degrees of freedom absorb excess entropy, facilitating the crystallization of $^3$He with increasing temperature.

Motivated by these advancements, one expects the situation to become more complex and intriguing if the system itself is a quantum many-body system. In this case, the interplay between intrinsic quantum fluctuations and the thermal fluctuations from the bath degrees of freedom may give rise to fascinating quantum phenomena that are absent in conventional high-temperature quantum systems. A key question is whether macroscopic quantum coherent states can be realized at arbitrarily high or even infinite temperatures. If so, what distinguishes these high-temperature quantum coherent states from their low-temperature counterparts?

\begin{figure}[htbp]
    \centering
    \includegraphics[width=0.45\textwidth]{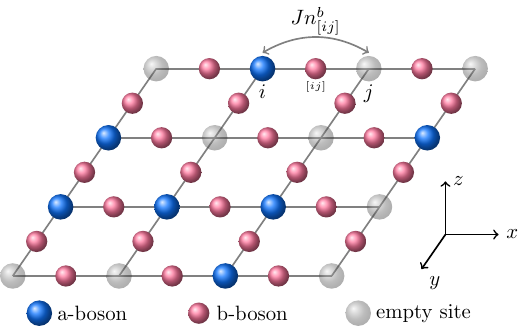}
    \caption{Schematic of the hybrid model composed of hard-core bosons (blue circles) and soft-core bosons (red circles), located on the lattice sites and bonds, respectively. The hopping amplitude of the hard-core bosons is mediated by the density of the bond bosons.}
    \label{fig:fig1}
\end{figure}

\begin{figure*}[htb]
    \includegraphics[width=1.0\textwidth]{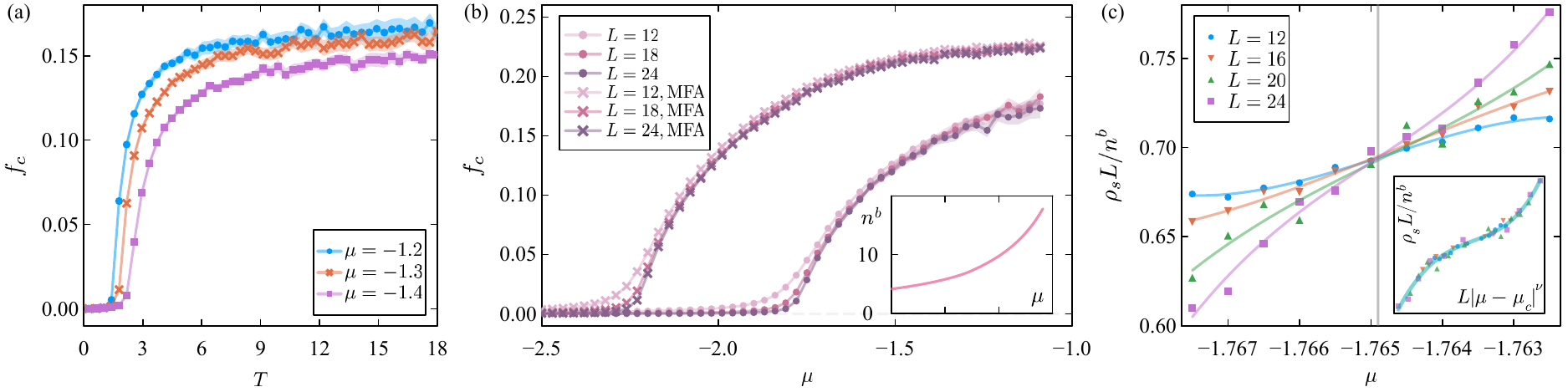}
    \caption{For the 3D model: (a) The condensate fraction of a-bosons, $f_c$, as a function of temperature for different values of $\mu$; (b) $f_c$ as a function of $\mu$, obtained from the QMC simulation (dotted curves) and the mean-field approximation (dashed curves). The inset shows the average density of the b-bosons as a function of $\mu$; (c) The normalized superfluid density of the a-bosons as a function of $\mu$ for different system sizes, which reveals a scaling-invariant critical point at $\mu_c \approx -1.764(9)$. The inset shows the data collapse for (c) using the critical exponent $\nu = 0.67$ (the same as for the 3D XY model). The temperature for (b) and (c) is fixed at $T = 10$.}
    \label{fig:fig2}
\end{figure*}

In this study, we propose a minimal model to address these issues. It is a lattice model composed of hard-core bosons (referred to as ``a-bosons"), whose hopping amplitude is mediated by the density of another type of soft-core bosons located on the lattice bonds (referred to as ``b-bosons"), as illustrated in {\FIG{fig:fig1}}. This hybrid bosonic model is free from the negative sign problem, making it amenable to solution through quantum Monte Carlo (QMC) simulations in a numerical precise way. We show that, due to their unbounded local degrees of freedom, the b-bosons can absorb entropy indefinitely, enabling the a-bosons to condense even at infinite temperature. Furthermore, it is possible to realize an infinite-temperature phase transition in a three-dimensional (3D) lattice by tuning the chemical potential of the b-bosons, which effectively adjusts the ``effective" temperature of the a-bosons within the mean-field approximation (MFA). However, in a two-dimensional (2D) lattice, the thermal fluctuations of the b-bosons play a critical role, causing the quasi-superfluid state predicted by the MFA to be completely suppressed. This gives way to a bosonic normal state, where the conductivity of the a-bosons is enhanced, rather than suppressed, by the fluctuations of the b-bosons.

\secheader{Model and method} We study a lattice model composed of two types of bosons, located on the lattice sites and bonds, respectively. The Hamiltonian of this hybrid system is given by:
\begin{equation}
\hat{H}(\{n_{[ij]}^b\}) = \sum_{[ij]} \left[ -J n_{[ij]}^b (\hat{a}_i^\dag \hat{a}_j + \mathrm{h.c.}) - \mu \hat{n}_{[ij]}^b \right], \label{eq:Ham}
\end{equation}
where the summation runs over all the bonds $[ij]$ (pairs of adjacent sites) on the lattice. $\hat{a}_i^\dag$ ($\hat{a}_i$) denotes the creation (annihilation) operator of the hard-core boson (a-boson), and $\hat{n}_{[ij]}^b$ is the density operator for the soft-core boson (b-boson), located on the bond $[ij]$. $J$ is the bare single-particle hopping amplitude for the a-bosons, which we set to be the unit of the energy scale throughout this study ($J=1$). $\mu$ is the chemical potential of the b-bosons. The key feature of this model is that the density operators of the b-bosons commute with the Hamiltonian, i.e., $[ \hat{n}_{[ij]}^b, \hat{H}(\{n_{[ij]}^b\}) ] = 0$. This implies that the Hilbert space can be divided into different subspaces based on the density configurations of the b-bosons, $\{n_{[ij]}^b\}$, with no quantum tunneling between these subspaces. Therefore, $\{n_{[ij]}^b\}$ can be treated as a set of ``classical" variables that label the different subspaces. The hybrid system is assumed to be in thermodynamic equilibrium, and its partition function reads:
\begin{equation}
Z = \sum_{\{n_{[ij]}^b\}} \mathrm{Tr} \, e^{-\beta \hat{H}(\{n_{[ij]}^b\})} , \label{eq:partition}
\end{equation}
where the summation runs over all possible density configurations of the b-bosons. Here, $\beta = \frac{1}{k_B T}$ is the inverse temperature, with $T$ being the temperature. Although the b-bosons themselves have no quantum dynamics, they couple to a heat bath that introduces thermal noise, accounting for the thermal fluctuations of $\{n_{[ij]}^b\}$.

To simulate such a hybrid system, we develop a QMC algorithm that allows us to simultaneously sample the world-line configurations of the a-bosons and the density configurations of the b-bosons, $\{n_{[ij]}^b\}$, according to their weights in the partition function (\ref{eq:partition}). For each given $\{n_{[ij]}^b\}$, the system reduces to a hard-core boson model with bond disorder, which can be simulated via the QMC algorithm with worm updates \cite{Prokofev1998,Pollet2005}. Since $n_{[ij]}^b \ge 0$, the b-bosons do not introduce a sign problem for the QMC simulation. The thermal fluctuations of the b-bosons necessitate an update of their density configurations $\{n_{[ij]}^b\}$. The details of the update algorithm and a proof of the detailed balance condition can be found in the supplementary material (SM) \cite{Supplementary}. 

We perform simulations on both 3D $L^3$ cubic lattices and 2D $L \times L$ square lattices with periodic boundary conditions. Under the MFA, the effective chemical potential of the b-bosons becomes $\mu_{\mathrm{eff}} = \mu + J \langle a_i^\dag a_j + a_j^\dag a_i \rangle$. Since the kinetic energy of the a-bosons satisfies $|\langle a_i^\dag a_j + a_j^\dag a_i \rangle| \leq 1$, and considering the soft-core nature of the b-bosons, the stability condition requires $\mu_{\mathrm{eff}} < 0$ ($\mu < -J$). If this condition is not met, the appearance of negative energy levels for the b-bosons causes $n_{[ij]}^b$ to diverge.

It is impossible to directly simulate the infinite-temperature case in the QMC algorithm. Therefore, we simulate a sufficiently high temperature at which {\it the physical quantities of the a-bosons have saturated to their values in the infinite-temperature limit.} An important advantage of our QMC algorithm, compared to other methods, is that there is no cutoff for the local occupation number of the b-bosons. As a result, the quantities of b-bosons never saturate. Since any realistic quantum system has a finite cutoff for the local Hilbert space dimension, the ``infinite" temperature regime discussed here corresponds to a temperature regime much higher than the typical energy scales of the a-bosons, while the average density of the b-bosons is still far from the cutoff of the local occupation number. A detailed discussion of the definition of the infinite-temperature limit can also be found in the SM \cite{Supplementary}.

\begin{figure}[htb]
    \includegraphics[width=0.45\textwidth]{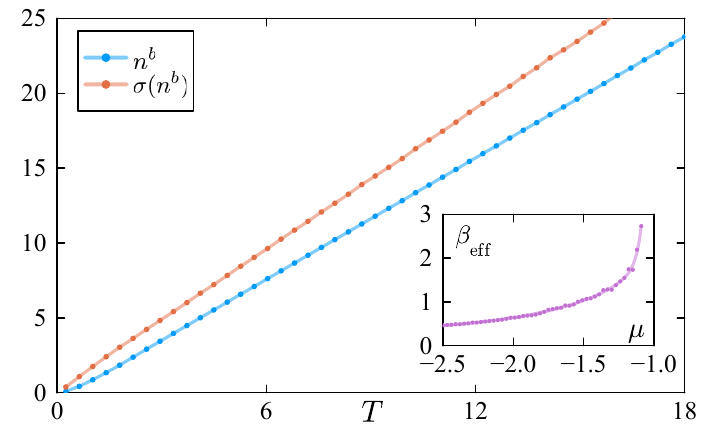}
    \caption{The average value $n^b$ (blue dot) and variance $\sigma(n^b)$ (red dot) of the density operators of the b-boson as a function of temperature. The parameters are chosen as $\mu = -1.3$ and $L = 16$ for a cubic lattice. The inset shows the effective inverse temperature $\beta_{\text{eff}}$ versus $\mu$ with a fixed $T = 10$.}
    \label{fig:fig3}
\end{figure}

\secheader{Infinite-temperature phases and phase transitions for 3D bosons} 
To begin with, we study the 3D case with lattice sites $N = L^3$, focusing on the condensate fraction:
\begin{equation}
f_c = \frac{1}{N^2} \sum_{ij} \langle \hat{a}_i^\dag \hat{a}_j \rangle,
\end{equation}
which characterizes the Bose-Einstein condensate (BEC) of the a-bosons. The summation $\sum_{ij}$ runs over all lattice sites, and $\langle \hat{O} \rangle = \frac{1}{Z} \operatorname{Tr} \left[ \hat{O} e^{-\beta \hat{H}} \right]$. We first fix $\mu$ and study the temperature dependence of $f_c$. As shown in {\FIG{fig:fig2}\,(a)}, in striking contrast to conventional BEC, our model exhibits an increase in $f_c$ with temperature $T$, which ultimately saturates to a finite value in the infinite-temperature limit ($T\rightarrow \infty$). Furthermore, for sufficiently high temperatures, where $f_c$ has saturated, tuning the chemical potential $\mu$ of the b-boson reveals a continuous phase transition characterized by the vanishing of the BEC of the a-bosons, as shown in {\FIG{fig:fig2}\,(b)}. 

To identify the universality class of this phase transition, we calculate the superfluid density $\rho_s$. In {\FIG{fig:fig2}\,(c)}, we plot the normalized $\rho_s L / n^b$ as a function of $\mu$ for different system sizes $L$, where $n^b = \langle \hat{n}^b_{[ij]} \rangle$ and $n^b J$ can be interpreted as the effective hopping amplitude for the a-bosons. {\FIG{fig:fig2}\,(c)} exhibits a crossing point, indicating a scaling-invariant critical point. To determine the critical exponents, we perform a data collapse analysis (see the inset of {\FIG{fig:fig2}\,(c)}), which yields the critical exponent $\nu = 0.67(1)$, in agreement with the critical exponent for the finite-temperature phase transition of 3D BEC.

These counterintuitive behaviors can be qualitatively understood within the MFA, where $\hat{n}^b_{[ij]}$ in \EQ{eq:Ham} is replaced by its average value, $n^b = \langle \hat{n}^b_{[ij]} \rangle$, neglecting thermal and spatial fluctuations. The temperature dependence of $n^b$ is shown in the inset of \FIG{fig:fig3}, where we observe that, in the high-temperature limit, $n^b$ grows linearly with temperature:
\begin{equation}
n^b = \beta_{\mathrm{eff}} T. \label{eq:nb}
\end{equation}
The slope $\beta_{\mathrm{eff}}$ depends on $\mu$, as shown in the inset of \FIG{fig:fig3}. $\beta_{\mathrm{eff}}$ diverges when $\mu \rightarrow -J$, corresponding to the zero-temperature limit in the MFA. The linear relation \EQ{eq:nb} can be understood by analyzing a single b-bosonic mode, where the backaction of the a-bosons is accounted for by an effective chemical potential $\mu_{\mathrm{eff}}$ in the MFA. Substituting \EQ{eq:nb} into \EQ{eq:partition}, we obtain $Z \sim \mathrm{Tr} \, e^{-\beta_{\mathrm{eff}} \hat{H}_a}$, where $\hat{H}_a$ represents the Hamiltonian of a conventional lattice model of hard-core bosons:
\begin{equation}
\hat{H}_a = \sum_{[ij]} -J (\hat{a}_i^\dag \hat{a}_j + \mathrm{h.c.}) .\label{eq:Ham2}
\end{equation}
Therefore, under the MFA, in the high-temperature limit, the divergence of $n^b$ compensates for the decrease in $\beta$, resulting in an effective temperature $T_{\mathrm{eff}} = \frac{1}{k_B \beta_{\mathrm{eff}}}$ for the a-bosons, which is significantly lower than the actual temperature $T$ ($T_{\mathrm{eff}}$ approaches a constant even as $T \to \infty$). The dependence of $T_{\mathrm{eff}}$ on $\mu$ also explains the infinite-temperature phase transition observed above, which corresponds to a finite ``effective temperature" phase transition of the Hamiltonian (\ref{eq:Ham2}).

\begin{figure*}[htb]
    \includegraphics[width=0.9\textwidth]{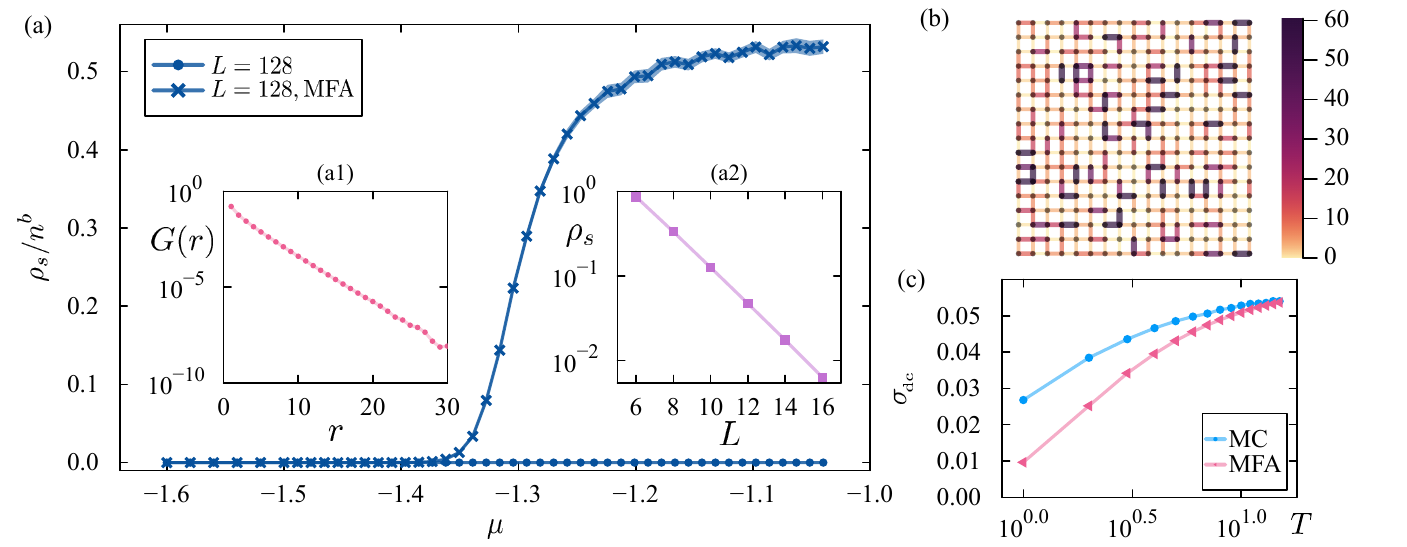}
    \caption{(a) For a 2D model, the normalized superfluid density of the a-bosons, $\rho_s / n_b$, as a function of $\mu$ obtained by the MFA (cross curves) and hybrid QMC (dot curves). The latter suggests the absence of a quasi-superfluid phase, which is further supported by the size dependence of $\rho_s$ (right inset) and the single-particle correlation function (left inset). (b) A typical snapshot of the $\{n^b_{[ij]}\}$ configuration generated during the hybrid QMC simulation. (c) A comparison of the DC conductivity of the a-bosons obtained by the MFA (blue circles) and hybrid QMC (red triangles). The parameters are chosen as $\mu = -1.2$, $T = 10$ for (a) and (b); $\mu = -2$ and $L = 16$ for (c).}
    \label{fig:fig4}
\end{figure*}

Despite its simplicity, the MFA ignores an important ingredient: the fluctuation of b-bosons. Taking the density operator of the b-boson on bond $[ij]$, $\hat{n}^b_{[ij]}$, as an example, we find that not only does its average value diverge, but also the fluctuation:
$\sigma^b = \sqrt{\langle (\hat{n}^b_{[ij]})^2 \rangle - \langle \hat{n}^b_{[ij]} \rangle^2}$
which linearly diverges with temperature in the high-temperature limit (see \FIG{fig:fig3}). The divergence of the fluctuation of $\hat{n}^b_{[ij]}$ invalidates the MFA in principle. Moreover, for a typical configuration of $\{n^b_{[ij]}\}$, the values of $n^b_{[ij]}$ vary significantly from bond to bond. As a result, the a-bosons hop in a lattice with strong bond disorder, which may significantly alter the physics of the a-bosons. To verify the effect of these fluctuations, we compare the $f_c - \mu$ curves obtained from the hybrid QMC simulation of the entire system with the results from the MFA treatment, as shown in \FIG{fig:fig2}\,(b). {\it To make this comparison meaningful, we first calculate $n^b$ from the hybrid QMC simulation of the Hamiltonian in \EQ{eq:Ham}, then substitute it into \EQ{eq:nb} to derive $\beta_{\mathrm{eff}}$. In the MFA, we perform the standard QMC simulation over the hard-core boson Hamiltonian (\ref{eq:Ham2}) under the effective temperature $T_{\mathrm{eff}}$ obtained from the hybrid QMC.} As shown in \FIG{fig:fig2}\,(b), the MFA overestimates the stability of the ordered phase, as expected, since fluctuations tend to suppress quantum coherence.

\secheader{Depletion of quasi-superfluid in square lattice: breakdown of mean-field approximation} Although fluctuations of b-bosons in a cubic lattice only quantitatively affect the phase diagram compared to the MFA predictions, they introduce a qualitative difference in a 2D lattice. For our hybrid model at sufficiently high temperatures, the MFA treatment leads to a 2D hard-core bosonic Hamiltonian (\ref{eq:Ham2}) with an effective temperature $T_{\mathrm{eff}}$ that depends on $\mu$. This model predicts a Berezinski-Kosterlitz-Thouless (BKT) transition\cite{Berezinskii1971, Kosterlitz1973} from a quasi-superfluid phase to a bosonic normal phase by tuning $\mu$ (see \FIG{fig:fig4}\,(a) for the MFA prediction). However, hybrid QMC simulations of the Hamiltonian in \EQ{eq:Ham} suggest the absence of a quasi-superfluid phase, even as $T_{\mathrm{eff}}$ approaches zero. In a finite system, the superfluid density $\rho_s$ of the a-bosons decays exponentially with the system size and vanishes in the thermodynamic limit, as shown in the right inset {\FIG{fig:fig4}\.(a2)}. Additionally, the correlation function of the a-bosons decays exponentially with distance, instead of algebraically (see the left inset {\FIG{fig:fig4}\,(a1)}), indicating a bosonic normal phase.

The absence of an infinite-temperature quasi-superfluid phase is a direct consequence of spatial fluctuations of the b-bosons. A typical snapshot of $\{n^b_{[ij]}\}$ generated by QMC sampling is shown in \FIG{fig:fig4}\,(b), where a rugged landscape for the a-bosons is clearly visible. While disorder typically suppresses superfluidity and can lead to a Bose glass phase\cite{Priyadarshee2006}, it is important to emphasize that the role of the b-boson density $\{n^b_{[ij]}\}$ in our model differs significantly from quenched bond disorder. The latter is static over time and tends to localize particles. In contrast, the b-bosons in our model are coupled to a heat bath, where thermal noises drastically changes $\{n^b_{[ij]}\}$ in time.

To assess the impact of these fluctuations on the transport properties of the a-bosons, we compute the conductivity of the a-bosons and compare it to the MFA result, where $\{n^b_{[ij]}\}$ is replaced by a uniform, static background ($n^b = \langle \hat{n}^b_{[ij]} \rangle$). In both cases, the DC conductivity $\sigma_{\mathrm{dc}}$ can be derived from the imaginary-time current-current correlation functions\cite{Scalettar1999} obtained via QMC simulations\cite{Supplementary}. The results for $\sigma_{\mathrm{dc}}$ obtained by the hybrid QMC and the MFA are plotted in \FIG{fig:fig4}\,(c). The comparison suggests that, although the fluctuations of $\{n^b_{[ij]}\}$ suppress the superfluidity of the a-bosons, they enhance their conductivity, making it larger than the conductivity predicted by the uniform MFA treatment without fluctuations.

\secheader{Discussion and generalization} 
We have proposed a mechanism for high-temperature macroscopic quantum coherent states. One may wonder whether a similar mechanism could give rise to other high-temperature quantum phases, such as superconductivity or non-Fermi liquid behavior. To explore this, we generalize our model to the fermionic case, using the Hamiltonian:
\begin{equation}
    \hat{H}_f = \sum_{[ij]} \left[ -J \hat{n}_{[ij]}^b \sum_{\sigma = \uparrow, \downarrow} \left( \hat{c}_{i\sigma}^\dag \hat{c}_{j\sigma} + \mathrm{h.c.} \right) - \mu \hat{n}_{[ij]}^b \right], \label{eq:Ham3}
 \end{equation}
where $\hat{c}_{i\sigma}$ ($\hat{c}^\dag_{i\sigma}$) is the annihilation (creation) operator for fermions with spin $\sigma$, and $\hat{n}_{[ij]}^b$ is defined as in Hamiltonian (\ref{eq:Ham}). Thermal averaging over the b-boson degrees of freedom will induce correlations between the fermions, thereby opening up an avenue to search for fermionic ordered states or even non-Fermi liquid phases. This quantum-classical hybrid model is reminiscent of the Falicov-Kimball model\cite{Falicov1969}, which consists of itinerant d-electrons and localized f-electrons with local interactions between them. In this model, the density of the f-electrons can be considered as a ``classical" degree of freedom. The model can be solved using Monte Carlo simulations and exhibits intriguing thermodynamic properties arising from the interplay between the classical and quantum degrees of  freedo \cite{Maska2006, Brandt1989, Brandt1990, Hohenadler2018}.

\secheader{Conclusion and outlook} 
In conclusion, we propose a quantum-classical hybrid model where the "classical bath" can absorb entropy indefinitely, leading to nontrivial quantum physics for the system at arbitrarily high temperatures. This model can be generalized to a fully quantum version by including terms that break the particle conservation of b-bosons (e.g., $(b^\dag_{[ij]} + b_{[ij]})$), thus treating b-bosons on the same footing as a-bosons. This approach is reminiscent of the Holstein model \cite{Holstein1959}. By performing QMC simulations \cite{Weber2016, Costa2018, Chen2019, Gotz2022}, the model can be solved numerically, raising the question of whether the nontrivial high-temperature quantum physics can persist in the presence of quantum fluctuations of the b-bosons. 

Another important direction for future work is the physical realization of high-temperature quantum coherent states. An oversimplified approach would be to couple the hard-core bosons to an optical cavity \cite{Baumann2010}, which provides a dynamically self-organized optical lattice whose depth varies with temperature, thus affecting the effective hopping amplitude of the hard-core bosons. This protocol corresponds to the MFA of our model with a uniform, temperature-dependent hopping amplitude. However, the inclusion of fluctuations remains an open question, deserving further exploration in future studies.

\begin{acknowledgments}
\secheader{Acknowledgments}
ZC  acknowledges helpful discussions with Xiaoyang Huang. This work is supported by the National Key Research and Development Program of China (2020YFA0309000, 2024YFA1408303), Natural Science Foundation of China (Grant \GrantNo{12174251} and \GrantNo{12525407}), Shanghai Municipal Science and Technology Major Project (Grant \GrantNo{2019SHZDZX01}), and Shanghai Science and Technology Innovation Action Plan (Grant \GrantNo{24Z510205936}). TZZ and ZZC are also supported by the National Center for High-Level Talent Training in Mathematics, Physics, Chemistry, and Biology.
\end{acknowledgments}


\begin{thebibliography}{35}
	\expandafter\ifx\csname natexlab\endcsname\relax\def\natexlab#1{#1}\fi
	\expandafter\ifx\csname bibnamefont\endcsname\relax
	\def\bibnamefont#1{#1}\fi
	\expandafter\ifx\csname bibfnamefont\endcsname\relax
	\def\bibfnamefont#1{#1}\fi
	\expandafter\ifx\csname citenamefont\endcsname\relax
	\def\citenamefont#1{#1}\fi
	\expandafter\ifx\csname url\endcsname\relax
	\def\url#1{\texttt{#1}}\fi
	\expandafter\ifx\csname urlprefix\endcsname\relax\def\urlprefix{URL }\fi
	\providecommand{\bibinfo}[2]{#2}
	\providecommand{\eprint}[2][]{\url{#2}}
	
	\bibitem[{\citenamefont{Pichler et~al.}(2010)\citenamefont{Pichler, Daley, and
			Zoller}}]{Pichler2010}
	\bibinfo{author}{\bibfnamefont{H.}~\bibnamefont{Pichler}},
	\bibinfo{author}{\bibfnamefont{A.~J.} \bibnamefont{Daley}}, \bibnamefont{and}
	\bibinfo{author}{\bibfnamefont{P.}~\bibnamefont{Zoller}},
	\bibinfo{journal}{Phys. Rev. A} \textbf{\bibinfo{volume}{82}},
	\bibinfo{pages}{063605} (\bibinfo{year}{2010}).
	
	\bibitem[{\citenamefont{Marino and Silva}(2012)}]{Marino2012}
	\bibinfo{author}{\bibfnamefont{J.}~\bibnamefont{Marino}} \bibnamefont{and}
	\bibinfo{author}{\bibfnamefont{A.}~\bibnamefont{Silva}},
	\bibinfo{journal}{Phys. Rev. B} \textbf{\bibinfo{volume}{86}},
	\bibinfo{pages}{060408} (\bibinfo{year}{2012}).
	
	\bibitem[{\citenamefont{Poletti et~al.}(2012)\citenamefont{Poletti, Bernier,
			Georges, and Kollath}}]{Poletti2012}
	\bibinfo{author}{\bibfnamefont{D.}~\bibnamefont{Poletti}},
	\bibinfo{author}{\bibfnamefont{J.-S.} \bibnamefont{Bernier}},
	\bibinfo{author}{\bibfnamefont{A.}~\bibnamefont{Georges}}, \bibnamefont{and}
	\bibinfo{author}{\bibfnamefont{C.}~\bibnamefont{Kollath}},
	\bibinfo{journal}{Phys. Rev. Lett.} \textbf{\bibinfo{volume}{109}},
	\bibinfo{pages}{045302} (\bibinfo{year}{2012}).
	
	\bibitem[{\citenamefont{Cai and Barthel}(2013)}]{Cai2013}
	\bibinfo{author}{\bibfnamefont{Z.}~\bibnamefont{Cai}} \bibnamefont{and}
	\bibinfo{author}{\bibfnamefont{T.}~\bibnamefont{Barthel}},
	\bibinfo{journal}{Phys. Rev. Lett.} \textbf{\bibinfo{volume}{111}},
	\bibinfo{pages}{150403} (\bibinfo{year}{2013}).
	
	\bibitem[{\citenamefont{Sieberer et~al.}(2013)\citenamefont{Sieberer, Huber,
			Altman, and Diehl}}]{Sieberer2013}
	\bibinfo{author}{\bibfnamefont{L.~M.} \bibnamefont{Sieberer}},
	\bibinfo{author}{\bibfnamefont{S.~D.} \bibnamefont{Huber}},
	\bibinfo{author}{\bibfnamefont{E.}~\bibnamefont{Altman}}, \bibnamefont{and}
	\bibinfo{author}{\bibfnamefont{S.}~\bibnamefont{Diehl}},
	\bibinfo{journal}{Phys. Rev. Lett.} \textbf{\bibinfo{volume}{110}},
	\bibinfo{pages}{195301} (\bibinfo{year}{2013}).
	
	\bibitem[{\citenamefont{Poletti et~al.}(2013)\citenamefont{Poletti, Barmettler,
			Georges, and Kollath}}]{Poletti2013}
	\bibinfo{author}{\bibfnamefont{D.}~\bibnamefont{Poletti}},
	\bibinfo{author}{\bibfnamefont{P.}~\bibnamefont{Barmettler}},
	\bibinfo{author}{\bibfnamefont{A.}~\bibnamefont{Georges}}, \bibnamefont{and}
	\bibinfo{author}{\bibfnamefont{C.}~\bibnamefont{Kollath}},
	\bibinfo{journal}{Phys. Rev. Lett.} \textbf{\bibinfo{volume}{111}},
	\bibinfo{pages}{195301} (\bibinfo{year}{2013}).
	
	\bibitem[{\citenamefont{Buchhold and Diehl}(2015)}]{Buchhold2015}
	\bibinfo{author}{\bibfnamefont{M.}~\bibnamefont{Buchhold}} \bibnamefont{and}
	\bibinfo{author}{\bibfnamefont{S.}~\bibnamefont{Diehl}},
	\bibinfo{journal}{Phys. Rev. A} \textbf{\bibinfo{volume}{92}},
	\bibinfo{pages}{013603} (\bibinfo{year}{2015}).
	
	\bibitem[{\citenamefont{Ren et~al.}(2020)\citenamefont{Ren, Li, Li, Cai, and
			Wang}}]{Cai2020}
	\bibinfo{author}{\bibfnamefont{J.}~\bibnamefont{Ren}},
	\bibinfo{author}{\bibfnamefont{Q.}~\bibnamefont{Li}},
	\bibinfo{author}{\bibfnamefont{W.}~\bibnamefont{Li}},
	\bibinfo{author}{\bibfnamefont{Z.}~\bibnamefont{Cai}}, \bibnamefont{and}
	\bibinfo{author}{\bibfnamefont{X.}~\bibnamefont{Wang}},
	\bibinfo{journal}{Phys. Rev. Lett.} \textbf{\bibinfo{volume}{124}},
	\bibinfo{pages}{130602} (\bibinfo{year}{2020}).
	
	\bibitem[{\citenamefont{Cai}(2022)}]{Cai2022}
	\bibinfo{author}{\bibfnamefont{Z.}~\bibnamefont{Cai}}, \bibinfo{journal}{Phys.
		Rev. Lett.} \textbf{\bibinfo{volume}{128}}, \bibinfo{pages}{050601}
	(\bibinfo{year}{2022}).
	
	\bibitem[{\citenamefont{Ljubotina et~al.}(2019)\citenamefont{Ljubotina,
			\ifmmode \check{Z}\else \v{Z}\fi{}nidari\ifmmode~\check{c}\else \v{c}\fi{},
			and Prosen}}]{Ljubotina2019}
	\bibinfo{author}{\bibfnamefont{M.}~\bibnamefont{Ljubotina}},
	\bibinfo{author}{\bibfnamefont{M.}~\bibnamefont{\ifmmode \check{Z}\else
			\v{Z}\fi{}nidari\ifmmode~\check{c}\else \v{c}\fi{}}}, \bibnamefont{and}
	\bibinfo{author}{\bibfnamefont{T.~c.~v.} \bibnamefont{Prosen}},
	\bibinfo{journal}{Phys. Rev. Lett.} \textbf{\bibinfo{volume}{122}},
	\bibinfo{pages}{210602} (\bibinfo{year}{2019}).
	
	\bibitem[{\citenamefont{De~Nardis et~al.}(2019)\citenamefont{De~Nardis,
			Medenjak, Karrasch, and Ilievski}}]{Nardis2019}
	\bibinfo{author}{\bibfnamefont{J.}~\bibnamefont{De~Nardis}},
	\bibinfo{author}{\bibfnamefont{M.}~\bibnamefont{Medenjak}},
	\bibinfo{author}{\bibfnamefont{C.}~\bibnamefont{Karrasch}}, \bibnamefont{and}
	\bibinfo{author}{\bibfnamefont{E.}~\bibnamefont{Ilievski}},
	\bibinfo{journal}{Phys. Rev. Lett.} \textbf{\bibinfo{volume}{123}},
	\bibinfo{pages}{186601} (\bibinfo{year}{2019}).
	
	\bibitem[{\citenamefont{Dupont and Moore}(2020)}]{Dupont2019}
	\bibinfo{author}{\bibfnamefont{M.}~\bibnamefont{Dupont}} \bibnamefont{and}
	\bibinfo{author}{\bibfnamefont{J.~E.} \bibnamefont{Moore}},
	\bibinfo{journal}{Phys. Rev. B} \textbf{\bibinfo{volume}{101}},
	\bibinfo{pages}{121106} (\bibinfo{year}{2020}).
	
	\bibitem[{\citenamefont{Weinberg}(1974)}]{Weinberg1974}
	\bibinfo{author}{\bibfnamefont{S.}~\bibnamefont{Weinberg}},
	\bibinfo{journal}{Phys. Rev. D} \textbf{\bibinfo{volume}{9}},
	\bibinfo{pages}{3357} (\bibinfo{year}{1974}).
	
	\bibitem[{\citenamefont{Chai et~al.}(2020{\natexlab{a}})\citenamefont{Chai,
			Chaudhuri, Choi, Komargodski, Rabinovici, and Smolkin}}]{Chai2020}
	\bibinfo{author}{\bibfnamefont{N.}~\bibnamefont{Chai}},
	\bibinfo{author}{\bibfnamefont{S.}~\bibnamefont{Chaudhuri}},
	\bibinfo{author}{\bibfnamefont{C.}~\bibnamefont{Choi}},
	\bibinfo{author}{\bibfnamefont{Z.}~\bibnamefont{Komargodski}},
	\bibinfo{author}{\bibfnamefont{E.}~\bibnamefont{Rabinovici}},
	\bibnamefont{and} \bibinfo{author}{\bibfnamefont{M.}~\bibnamefont{Smolkin}},
	\bibinfo{journal}{Phys. Rev. Lett.} \textbf{\bibinfo{volume}{125}},
	\bibinfo{pages}{131603} (\bibinfo{year}{2020}{\natexlab{a}}).
	
	\bibitem[{\citenamefont{Chai et~al.}(2020{\natexlab{b}})\citenamefont{Chai,
			Chaudhuri, Choi, Komargodski, Rabinovici, and Smolkin}}]{Chai2020b}
	\bibinfo{author}{\bibfnamefont{N.}~\bibnamefont{Chai}},
	\bibinfo{author}{\bibfnamefont{S.}~\bibnamefont{Chaudhuri}},
	\bibinfo{author}{\bibfnamefont{C.}~\bibnamefont{Choi}},
	\bibinfo{author}{\bibfnamefont{Z.}~\bibnamefont{Komargodski}},
	\bibinfo{author}{\bibfnamefont{E.}~\bibnamefont{Rabinovici}},
	\bibnamefont{and} \bibinfo{author}{\bibfnamefont{M.}~\bibnamefont{Smolkin}},
	\bibinfo{journal}{Phys. Rev. D} \textbf{\bibinfo{volume}{102}},
	\bibinfo{pages}{065014} (\bibinfo{year}{2020}{\natexlab{b}}).
	
	\bibitem[{\citenamefont{Han et~al.}(2025)\citenamefont{Han, Huang, Komargodski,
			Lucas, and Popov}}]{Han2025}
	\bibinfo{author}{\bibfnamefont{Y.}~\bibnamefont{Han}},
	\bibinfo{author}{\bibfnamefont{X.}~\bibnamefont{Huang}},
	\bibinfo{author}{\bibfnamefont{Z.}~\bibnamefont{Komargodski}},
	\bibinfo{author}{\bibfnamefont{A.}~\bibnamefont{Lucas}}, \bibnamefont{and}
	\bibinfo{author}{\bibfnamefont{F.~K.} \bibnamefont{Popov}},
	\emph{\bibinfo{title}{Entropic order}} (\bibinfo{year}{2025}),
	\eprint{2503.22789}.
	
	\bibitem[{\citenamefont{Pomeranchuk}(1950)}]{Pomeranchuk1950}
	\bibinfo{author}{\bibfnamefont{I.}~\bibnamefont{Pomeranchuk}},
	\bibinfo{journal}{JETP} \textbf{\bibinfo{volume}{20}}, \bibinfo{pages}{919}
	(\bibinfo{year}{1950}).
	
	\bibitem[{\citenamefont{Prokof'ev et~al.}(1998)\citenamefont{Prokof'ev,
			Svistunov, and Tupitsyn}}]{Prokofev1998}
	\bibinfo{author}{\bibfnamefont{N.~V.} \bibnamefont{Prokof'ev}},
	\bibinfo{author}{\bibfnamefont{B.~V.} \bibnamefont{Svistunov}},
	\bibnamefont{and} \bibinfo{author}{\bibfnamefont{I.~S.}
		\bibnamefont{Tupitsyn}}, \bibinfo{journal}{Phys. Lett. A}
	\textbf{\bibinfo{volume}{238}}, \bibinfo{pages}{253} (\bibinfo{year}{1998}).
	
	\bibitem[{\citenamefont{Pollet et~al.}(2007)\citenamefont{Pollet, Houcke, and
			Rombouts}}]{Pollet2005}
	\bibinfo{author}{\bibfnamefont{L.}~\bibnamefont{Pollet}},
	\bibinfo{author}{\bibfnamefont{K.~V.} \bibnamefont{Houcke}},
	\bibnamefont{and} \bibinfo{author}{\bibfnamefont{S.~M.~A.}
		\bibnamefont{Rombouts}}, \bibinfo{journal}{J. Comp. Phys}
	\textbf{\bibinfo{volume}{225}}, \bibinfo{pages}{2249} (\bibinfo{year}{2007}).
	
	\bibitem[{Sup()}]{Supplementary}
	\bibinfo{howpublished}{See the supplementary material for details of the
		numerical simulation and a discussion on the definition of the infinite
		temperature limit.}
	
	\bibitem[{\citenamefont{Berezinskii}(1971)}]{Berezinskii1971}
	\bibinfo{author}{\bibfnamefont{V.~L.} \bibnamefont{Berezinskii}},
	\bibinfo{journal}{Sov. Phys. JETP} \textbf{\bibinfo{volume}{32}},
	\bibinfo{pages}{493} (\bibinfo{year}{1971}).
	
	\bibitem[{\citenamefont{Kosterlitz and Thouless}(1973)}]{Kosterlitz1973}
	\bibinfo{author}{\bibfnamefont{J.~M.} \bibnamefont{Kosterlitz}}
	\bibnamefont{and} \bibinfo{author}{\bibfnamefont{D.~J.}
		\bibnamefont{Thouless}}, \bibinfo{journal}{Journal of Physics C: Solid State
		Physics} \textbf{\bibinfo{volume}{6}}, \bibinfo{pages}{1181}
	(\bibinfo{year}{1973}).
	
	\bibitem[{\citenamefont{Priyadarshee et~al.}(2006)\citenamefont{Priyadarshee,
			Chandrasekharan, Lee, and Baranger}}]{Priyadarshee2006}
	\bibinfo{author}{\bibfnamefont{A.}~\bibnamefont{Priyadarshee}},
	\bibinfo{author}{\bibfnamefont{S.}~\bibnamefont{Chandrasekharan}},
	\bibinfo{author}{\bibfnamefont{J.-W.} \bibnamefont{Lee}}, \bibnamefont{and}
	\bibinfo{author}{\bibfnamefont{H.~U.} \bibnamefont{Baranger}},
	\bibinfo{journal}{Phys. Rev. Lett.} \textbf{\bibinfo{volume}{97}},
	\bibinfo{pages}{115703} (\bibinfo{year}{2006}).
	
	\bibitem[{\citenamefont{Scalettar et~al.}(1999)\citenamefont{Scalettar,
			Trivedi, and Huscroft}}]{Scalettar1999}
	\bibinfo{author}{\bibfnamefont{R.~T.} \bibnamefont{Scalettar}},
	\bibinfo{author}{\bibfnamefont{N.}~\bibnamefont{Trivedi}}, \bibnamefont{and}
	\bibinfo{author}{\bibfnamefont{C.}~\bibnamefont{Huscroft}},
	\bibinfo{journal}{Phys. Rev. B} \textbf{\bibinfo{volume}{59}},
	\bibinfo{pages}{4364} (\bibinfo{year}{1999}).
	
	\bibitem[{\citenamefont{Falicov and Kimball}(1969)}]{Falicov1969}
	\bibinfo{author}{\bibfnamefont{L.~M.} \bibnamefont{Falicov}} \bibnamefont{and}
	\bibinfo{author}{\bibfnamefont{J.~C.} \bibnamefont{Kimball}},
	\bibinfo{journal}{Phys. Rev. Lett.} \textbf{\bibinfo{volume}{22}},
	\bibinfo{pages}{997} (\bibinfo{year}{1969}).
	
	\bibitem[{\citenamefont{Ma\ifmmode~\acute{s}\else \'{s}\fi{}ka and
			Czajka}(2006)}]{Maska2006}
	\bibinfo{author}{\bibfnamefont{M.~M.} \bibnamefont{Ma\ifmmode~\acute{s}\else
			\'{s}\fi{}ka}} \bibnamefont{and}
	\bibinfo{author}{\bibfnamefont{K.}~\bibnamefont{Czajka}},
	\bibinfo{journal}{Phys. Rev. B} \textbf{\bibinfo{volume}{74}},
	\bibinfo{pages}{035109} (\bibinfo{year}{2006}).
	
	\bibitem[{\citenamefont{Brandt and Mielsch}(1989)}]{Brandt1989}
	\bibinfo{author}{\bibfnamefont{U.}~\bibnamefont{Brandt}} \bibnamefont{and}
	\bibinfo{author}{\bibfnamefont{C.}~\bibnamefont{Mielsch}},
	\bibinfo{journal}{Z. Phys. B - Condensed Matter}
	\textbf{\bibinfo{volume}{75}}, \bibinfo{pages}{365} (\bibinfo{year}{1989}).
	
	\bibitem[{\citenamefont{Brandt and Mielsch}(1990)}]{Brandt1990}
	\bibinfo{author}{\bibfnamefont{U.}~\bibnamefont{Brandt}} \bibnamefont{and}
	\bibinfo{author}{\bibfnamefont{C.}~\bibnamefont{Mielsch}},
	\bibinfo{journal}{Z. Phys. B - Condensed Matter}
	\textbf{\bibinfo{volume}{79}}, \bibinfo{pages}{295} (\bibinfo{year}{1990}).
	
	\bibitem[{\citenamefont{Hohenadler and Assaad}(2018)}]{Hohenadler2018}
	\bibinfo{author}{\bibfnamefont{M.}~\bibnamefont{Hohenadler}} \bibnamefont{and}
	\bibinfo{author}{\bibfnamefont{F.~F.} \bibnamefont{Assaad}},
	\bibinfo{journal}{Phys. Rev. Lett.} \textbf{\bibinfo{volume}{121}},
	\bibinfo{pages}{086601} (\bibinfo{year}{2018}).
	
	\bibitem[{\citenamefont{Holstein}(1959)}]{Holstein1959}
	\bibinfo{author}{\bibfnamefont{J.}~\bibnamefont{Holstein}},
	\bibinfo{journal}{Ann. Phys.(N.Y.)} \textbf{\bibinfo{volume}{8}},
	\bibinfo{pages}{325} (\bibinfo{year}{1959}).
	
	\bibitem[{\citenamefont{Weber et~al.}(2016)\citenamefont{Weber, Assaad, and
			Hohenadler}}]{Weber2016}
	\bibinfo{author}{\bibfnamefont{M.}~\bibnamefont{Weber}},
	\bibinfo{author}{\bibfnamefont{F.~F.} \bibnamefont{Assaad}},
	\bibnamefont{and}
	\bibinfo{author}{\bibfnamefont{M.}~\bibnamefont{Hohenadler}},
	\bibinfo{journal}{Phys. Rev. B} \textbf{\bibinfo{volume}{94}},
	\bibinfo{pages}{245138} (\bibinfo{year}{2016}).
	
	\bibitem[{\citenamefont{Costa et~al.}(2018)\citenamefont{Costa, Blommel, Chiu,
			Batrouni, and Scalettar}}]{Costa2018}
	\bibinfo{author}{\bibfnamefont{N.~C.} \bibnamefont{Costa}},
	\bibinfo{author}{\bibfnamefont{T.}~\bibnamefont{Blommel}},
	\bibinfo{author}{\bibfnamefont{W.-T.} \bibnamefont{Chiu}},
	\bibinfo{author}{\bibfnamefont{G.}~\bibnamefont{Batrouni}}, \bibnamefont{and}
	\bibinfo{author}{\bibfnamefont{R.~T.} \bibnamefont{Scalettar}},
	\bibinfo{journal}{Phys. Rev. Lett.} \textbf{\bibinfo{volume}{120}},
	\bibinfo{pages}{187003} (\bibinfo{year}{2018}).
	
	\bibitem[{\citenamefont{Chen et~al.}(2019)\citenamefont{Chen, Xu, Meng, and
			Hohenadler}}]{Chen2019}
	\bibinfo{author}{\bibfnamefont{C.}~\bibnamefont{Chen}},
	\bibinfo{author}{\bibfnamefont{X.~Y.} \bibnamefont{Xu}},
	\bibinfo{author}{\bibfnamefont{Z.~Y.} \bibnamefont{Meng}}, \bibnamefont{and}
	\bibinfo{author}{\bibfnamefont{M.}~\bibnamefont{Hohenadler}},
	\bibinfo{journal}{Phys. Rev. Lett.} \textbf{\bibinfo{volume}{122}},
	\bibinfo{pages}{077601} (\bibinfo{year}{2019}).
	
	\bibitem[{\citenamefont{G\"otz et~al.}(2022)\citenamefont{G\"otz, Beyl,
			Hohenadler, and Assaad}}]{Gotz2022}
	\bibinfo{author}{\bibfnamefont{A.}~\bibnamefont{G\"otz}},
	\bibinfo{author}{\bibfnamefont{S.}~\bibnamefont{Beyl}},
	\bibinfo{author}{\bibfnamefont{M.}~\bibnamefont{Hohenadler}},
	\bibnamefont{and} \bibinfo{author}{\bibfnamefont{F.~F.}
		\bibnamefont{Assaad}}, \bibinfo{journal}{Phys. Rev. B}
	\textbf{\bibinfo{volume}{105}}, \bibinfo{pages}{085151}
	(\bibinfo{year}{2022}).
	
	\bibitem[{\citenamefont{Baumann et~al.}(2010)\citenamefont{Baumann, Guerlin,
			Brennecke, and Esslinger}}]{Baumann2010}
	\bibinfo{author}{\bibfnamefont{K.}~\bibnamefont{Baumann}},
	\bibinfo{author}{\bibfnamefont{C.}~\bibnamefont{Guerlin}},
	\bibinfo{author}{\bibfnamefont{F.}~\bibnamefont{Brennecke}},
	\bibnamefont{and}
	\bibinfo{author}{\bibfnamefont{T.}~\bibnamefont{Esslinger}},
	\bibinfo{journal}{Nature} \textbf{\bibinfo{volume}{464}},
	\bibinfo{pages}{1301} (\bibinfo{year}{2010}).
	
\end{thebibliography}

\end{document}